\def\bey{\begin{eqnarray}}
\def\eey{\end{eqnarray}}
\def\be{\begin{equation}}
\def\ee{\end{equation}}
\def\ba{\begin{array}}
\def\ea{\end{array}}
\def\gm{\gamma}

\def\Ld{\Lambda}
\def\af{\alpha}
\def\sg{\sigma}

\def\om{\omega}
\def\r{\rho}

\def\bt{\beta}

\def\dt{\delta}

\def\pp{\partial}
\def\pp{\partial}

\def\nnb{\nonumber}
\bigskip
\documentclass[onecolumn,showpacs,preprintnumbers,amsmath,amssymb]{revtex4}
\usepackage{graphicx}
\usepackage{fancyhdr}
\usepackage{dcolumn}
\usepackage{bm}
\begin{document}
\preprint{ }
\title{ Effects of the density dependence of nuclear symmetry energy on
properties of superheavy nuclei}
\author{Wei-Zhou Jiang}
\affiliation{  $^1$ Department of Physics, Southeast University,
Nanjing 211189, China\\ $^2$ National  Laboratory of Heavy Ion
Accelerator, Lanzhou 730000,China}

\begin{abstract}
\baselineskip18pt Effects of the density dependence of the nuclear
symmetry energy on ground-state properties of superheavy nuclei are
studied in the relativistic mean-field theory. It is found that the
softening of the symmetry energy plays an important role in the
empirical shift [Phys. Rev. C \textbf{67}, 024309 (2003)] of
spherical orbitals in superheavy nuclei. The calculation based on the
relativistic mean-field models NL3 and FSUGold supports the double
shell closure in $^{292}120$ with the softening of the symmetry
energy.  In addition, the  significant effect of the density
dependence of the symmetry energy on the neutron skin thickness in
superheavy nuclei are investigated.
\end{abstract}
\maketitle \baselineskip 20.6pt

\section{Introduction}
The persistent interest in the synthesis of superheavy nuclei (SHN)
acquires refreshment due to recent progresses
~\cite{hof98,hof00,wi00,og04,og041,mo04,gan04,dv06,og06,og07,og071,
so07,ne08,mo08}. This is a hot field where people often expect the
next new superheavy element (SHE) that can be synthesized in the
laboratory. Indeed, since the cross section of the SHE synthesis is
very small,  it is much more difficult to synthesize the heavier and
heavier SHE~\cite{hof00}. For instance, the cross section of the cold
fusion reduces almost exponentially with the increase of the nuclear
charge in the superheavy region. One of important factors that
affects the synthesis is the shell closure in superheavy nuclei.
However, predictions turned out to be quite divisive for various
theoretical approaches. For instance, the microscopic-macroscopic
model predicts the the double shell closure at
(Z=114,N=184)~\cite{mol94}; various nonrelativistic models with
Skyrme forces can predict different double shell closures at
(Z=114,N=184)~\cite{ben99}, (120,172)~\cite{ben99,rut97,dec03} or
(126,184)~\cite{ben99,rut97,cwi96}, while most relativistic
mean-field (RMF) models incline the double shell closure at
(120,172)~\cite{ben99,rut97,zh05}. In general, the diversity of
predictions on the shell closure in the superheavy region is
associated with various single-particle properties near the Fermi
surface.

In the recent decade, the extraction of the constraint on the density
dependence of the symmetry energy has been another hot spot in
nuclear physics due to the availability of  high-quality radioactive
beams. The density dependence of the symmetry energy plays important
roles in understanding many important issues in astrophysics, see,
e.g., Refs.~\cite{lat01,ho01,st05}, properties of proton- or
neutron-rich nuclei and the reaction dynamics of heavy-ion
collisions, see, e.g., Refs. ~\cite{todd,ji05,ba02,ba08}. However,
the density dependence of the symmetry energy is still poorly known
especially at high densities~\cite{ba08}. Recent extraction of the
neutron skin thickness of $^{208}$Pb from collective flow data of
heavy ion collisions~\cite{ts04,ch05,li05} exhibited the softening
tendency of the symmetry energy. Since the density dependence of the
symmetry energy can reflect the surface property of the isovector
potential, the effect on the single-particle property and
characteristic of the shell closure in SHN may be induced by the
softening of the symmetry energy. Moreover, it was found that the
existence of the central depression is important for the double shell
closure in $^{292}120$~\cite{ben99,de99,af05}. In presence of the
central depression, the sensitivity of the  properties of SHN to
various density dependences of the symmetry energy can be affected.
Though the properties of SHN have been explored in a great number of
works~\cite{mol94,ben99,rut97,dec03,cwi96,zh05,de99,af03,af05,af06,ren01,
ren02, ren021,ren03,typ03,xu04,sil04,bar05,sha05,pei07,don08}, the
investigation on the symmetry energy dependent effect  is scarce.
Thus, it is the aim of this work to investigate the effect of the
softening of the symmetry energy on ground-state properties of SHN,
especially the shell closure.

In the past, the isoscalar-isovector coupling was first introduced in
RMF models to mimic various density dependences of the symmetry
energy in Ref.~\cite{ho01}, and its effects on the properties of
finite nuclei, nuclear matter and neutron stars have extensively been
investigated in the
literature~\cite{ho01,todd,ji05,ho01a,ho02,fsu,sh05,ji06}. It is an
economic way to simulate various density dependences of the symmetry
energy with the inclusion of the isoscalar-isovector coupling in RMF
models. In addition, the RMF theory is successful in describing the
properties of nuclei from proton drip line to neutron drip line,
since it can provide a dynamic description for the spin-orbit
interaction, e.g., see reviews in Refs.~\cite{se86,ga90,ri96}. In
this work, we thus perform the investigation with RMF models. The
paper is arranged in the following. In section~\ref{rmf}, the brief
formulas are given for RMF models. The results and discussions are
presented in section~\ref{results}. At last, a summary is given in
section~\ref{summary}.

\section{A brief formalism}
\label{rmf}
 The relativistic lagrangian can be written as:
 \bey
 {\cal L}&=&
{\overline\psi}[i\gm_{\mu}\partial^{\mu}-M_N+g_{\sg}\sg-g_{\om}
\gm_{\mu}\om^{\mu}-g_\r\gm_\mu \tau_3 b_0^\mu
  -e\frac{1}{2}(1+\tau_3)\gm_\mu A^\mu]\psi\nnb\\
      &  &
    - \frac{1}{4}F_{\mu\nu}F^{\mu\nu}+
      \frac{1}{2}m_{\om}^{2}\om_{\mu}\om^{\mu}
    - \frac{1}{4}B_{\mu\nu} B^{\mu\nu}+
      \frac{1}{2}m_{\r}^{2} b_{0\mu} b_0^{\mu}-\frac{1}{4}A_{\mu\nu}
      A^{\mu\nu}\nnb\\
&& +
\frac{1}{2}(\partial_{\mu}\sg\partial^{\mu}\sg-m_{\sg}^{2}\sg^{2})
+U(\sg,\om^\mu, b_0^\mu), \label{eq:lag1}
  \eey
 where $\psi,\sg,\om$, and $b_0$ are the fields of
the nucleon,  scalar, vector, and neutral isovector-vector,  with
their masses $M_N, m_\sg,m_\om$, and $m_\r$, respectively. $A_\mu$ is
the photon field. $g_i(i=\sg,\om,\r)$  are the corresponding
meson-nucleon couplings. $F_{\mu\nu}$, $ B_{\mu\nu}$ and $A_{\mu\nu}$
are the strength tensors of $\om$ and $\r$ mesons, and photon,
respectively
\begin{equation}\label{strength} F_{\mu\nu}=\pp_\mu
\om_\nu -\pp_\nu \om_\mu,\hbox{  } B_{\mu\nu}=\pp_\mu b_{0\nu}
-\pp_\nu b_{0\mu},\hbox{ } A_{\mu\nu}=\pp_\mu A_{\nu} -\pp_\nu
A_{\mu}.
\end{equation}
The self-interacting terms of $\sigma$, $\om$ mesons and  the
isoscalar-isovector coupling  are given generally as
 \bey
 U(\sg,\om^\mu, b_0^\mu)&=&-\frac{1}{3}g_2\sg^3-\frac{1}{4}g_3\sg^4
 +\frac{1}{4}c_3(\om_\mu\om^\mu)^2\nnb\\
 &&+4g^2_\r  g_\om^2 \Ld_{\rm v}
 \om_\mu\om^\mu b_{0\mu}b_0^\mu.
 \eey
Here, the isoscalar-isovector coupling term is introduced to modify
the density dependence of the symmetry energy.

Using the Euler-Lagrangian equation, the equations of motion for
nucleons and mesons can be obtained. In the RMF approximation, the
mesons are approximated by their  classic fields with quantum motion
neglected. The Dirac equation in RMF is written as
 \be [-i{\bf \af}\cdot\nabla +\bt M^*_N+
g_\om\om_0(r)+g_\r\tau_3 b_0(r)+e\frac{1}{2}(1+\tau_3)
A_0(r)]\psi_\af(r)=E_\af\psi_\af, \label{eqr} \ee with $M^*_N=M_N-
g_\sg\sg(r) $ and  $E_\af$ the single-particle energy. For
simplicity, the isospin subscript for the $\r$-meson field is omitted
hereafter. For the mesons and photon, the equations of motion are
given as
 \be
  (\Delta-m_\phi^2)\phi(r) =
  -s_\phi(r)
  \ee
where for the photon, $m_\phi=0$, and
 \be s_\phi(r)=\left\{\ba{cl}
g_\sg\r_s(r)-g_2\sg^2(r)-g_3\sg^3(r), &
 \hbox{ $\sg$ },\\\\
 g_\om\r_B(r)-c_3\om_0^3-8g^2_\om g^2_\r\Ld_{\rm v}\om_0(r) b_0^2(r), &
 \hbox{ $\om$ },\\\\
g_\r\r_3(r)-8g^2_\r g^2_\om\Ld_{\rm v} b_0(r) \om_0^2(r), &
 \hbox{ rho },\\\\
  e\r_c(r),&
\hbox{ photon.}
\\ \ea\right.
 \ee
Here $\r_s$, $\r_B$, $\r_3$ and $\r_c$ are the scalar, vector,
isovector and charge densities, respectively. We see that the fields
$b_0$ and $\om_0$ can be modified by the isoscalar-isovector
coupling. This modification can also affect the spin-orbit potential
which is written as \be
U_{ls}=\frac{1}{2M_\epsilon^2}\frac{d}{rdr}(V^\sg(r)+V(r)){\bf
L}\cdot{\bf S}, \ee
 where \bey\label{eqpot}
 M_\epsilon&=&M_N^*+E_\af-V(r)\approx
2M_N-(g_\sg\sg(r)+V(r)),\nnb\\
V(r)&=&g_\om \om_0(r)+g_\r b_0(r)t+e(\frac{1+t}{2})A_0,
  \eey
with  $t=\pm1$ for the proton and neutron, respectively.

The total binding energy is given as
 \bey \label{eqeng}
B&=&E_{N}+E_\sg+E_{\om_0}+E_{b_0}+E_c
+E_{CM}\nnb\\
 &=&\sum_\af (E_\af-M_N) -\frac{1}{2}\int d^3r [g_\sg\sg(r)\r_s(r)+
 \frac{1}{3}g_2\sg^3(r)+\frac{1}{2}g_3\sg^4(r) ]\nnb\\
 && +
 \frac{1}{2}\int
 d^3r[g_\om\om_0(r)\r_B(r)+\frac{c_3}{2}\om_0^4(r)]
 \nnb\\
&&+ \frac{1}{2}g_\r\int d^3rb_0(r)[\r_3(r)+8g_\r g_\om^2\Ld_{\rm v}
 \om_0^2(r)b_0(r)]  \nnb\\
  &&+ \frac{1}{2} e\int d^3rA_0(r)\r_c(r)-\frac{3}{4}41A^{1/3}.
 \eey

In practical calculations, the BCS pairing interaction is also
included using the constant pairing gaps which are obtained from the
prescription of M\"oller and Nix \cite{mn2}:
$\Delta_n=4.8/N^{1/3},\hbox{ } \Delta_p=4.8/Z^{1/3}$ with N and Z the
neutron and proton numbers, respectively. This prescription was also
used for the SHN in Ref.~\cite{sha05}. The cut-off $82A^{-1/3}$ MeV
above the nucleon chemical potentials is used to normalize the
pairing energy~\cite{ga90}. The coupled Dirac and meson equations are
solved for spherical nuclei with an iterative procedure. Details in
solving the equations can be found in the
literature~\cite{ga90,se86,ri96}, and are not reiterated here.

\section{Results and discussions}
\label{results}

\begin{figure}[thb]
\begin{center}
\vspace*{-25mm}
\includegraphics[height=14.0cm,width=12.0cm]{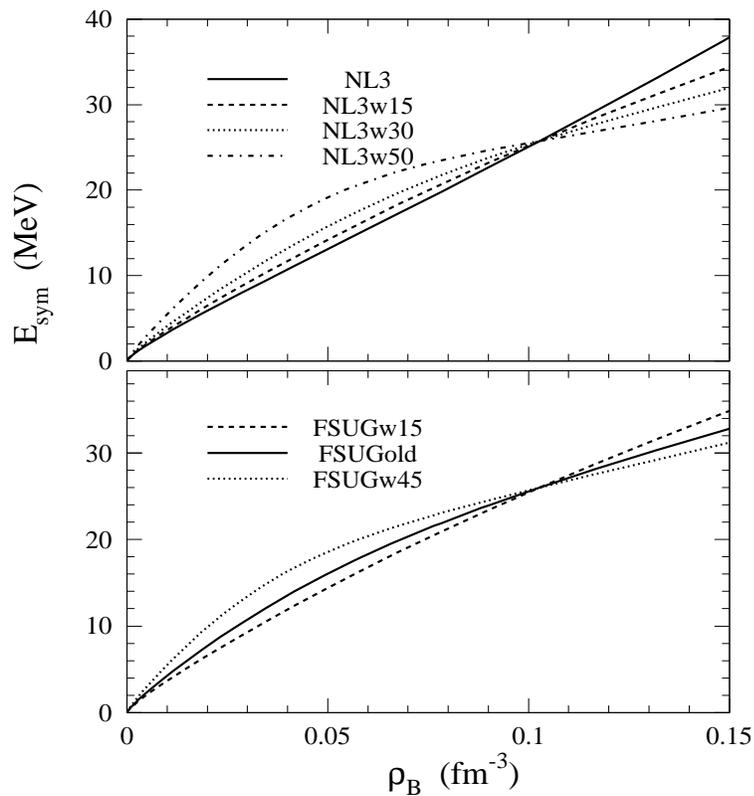}
 \end{center}
\vspace*{-10mm}\caption{Density dependence of the symmetry energy
with various isoscalar-isovector couplings in NL3 and
FSUGold.\label{figa4}}
\end{figure}

We first study the properties of the SHN with the RMF parameter set
NL3~\cite{nl3} where the isoscalar-isovector coupling is taken into
account to mimic various density dependences of the symmetry energy.
For comparisons, calculations are also performed with the RMF
parameter set FSUGold~\cite{fsu} that features the
isoscalar-isovector coupling. In RMF models, the symmetry energy can
be written as:
\begin{equation}\label{eqsym}
    E_{sym}=\frac{1}{2}\left(\frac{g_\r}{m_\r^*}\right)^2 \r_B
    +\frac{k_F^2}{6E_F^*}=\frac{1}{2\dt}g_\r b_0
    +\frac{k_F^2}{6E_F^*},
\end{equation}
where $m_\r^*$ is the $\r$-meson effective mass with
$m^*_\r=\sqrt{m_\r^2+8\Ld_{\rm v}(g_\om g_\r\om_0)^2}$, $\dt$ is the
isospin asymmetry  with $\dt=\r_3/\r_B$, and $E_F^*$ is the Fermi
energy. The first term is the potential part of the symmetry energy,
and the second term is the kinetic part. The modification to the
symmetry energy is dictated by the potential part through the
isoscalar-isovector coupling.  For a given $\Ld_{\rm v}$, we follow
Ref.~\cite{ho01} to readjust the $\r NN$ coupling constant $g_\r$ so
as to keep the symmetry energy unchanged at $k_F=1.15$ fm$^{-1}$
($\r=0.7\r_0$). As shown in Fig.~\ref{figa4}, the symmetry energy is
softened by the isoscalar-isovector coupling.   With this softening
of the symmetry energy, the appreciable reduction of the neutron skin
thickness in heavy nuclei can be obtained without compromising the
success in reproducing a variety of ground-state
properties~\cite{ho01}. Due to the inclusion of the
isoscalar-isovector coupling, the total binding energy of heavy
nuclei changes by a few MeV, and in SHN this change can rise
moderately. To reduce the variation of the binding energy in SHN, one
may readjust slightly the parameters such as the meson-nucleon
coupling constants and mesons. Without priority, here we readjust
slightly the $\sg$ meson mass $m_\sg$. For simplicity, we do not
perform the best-fit procedure, and the value of $m_\sg$ is just
refitted to the binding energy of $^{208}$Pb. The readjusted
parameters with various $\Ld_{\rm v}$ and properties of $^{208}$Pb
are listed in table~\ref{tab1}. Except for the original parameter
sets NL3 and FSUGold,  other parameter sets listed in
Table~\ref{tab1} are named according to the value of $\Ld_{\rm v}$.
Next, we perform calculations for SHN with these parameter sets and
examine the sensitivity of ground-state properties of SHN to
differences in the symmetry energy.

 \begin{table}[bht]
\caption{Readjusted parameters in NL3 and FSUGold with ground-state
properties of $^{208}$Pb.  The binding energy per nucleon (B/A),
proton radius ($r_p$) and neutron skin thickness ($r_p-r_n$) are
listed. The slightly modified incompressibility is listed in the last
column.  } \label{tab1}
 \begin{center}
    \begin{tabular}{ c| c c c c c c c}
\hline
 Model&$\Ld_{\rm v}$&$g_\r$ &$m_\sg$ (MeV) & B/A (MeV) & $r_p$ (fm) & $r_n-r_p$
 (fm) & $\kappa$ (MeV)\\ \hline
NL3    & 0.000  &  4.4740 & 508.194  & 7.889&   5.459   &  0.281&271.78   \\
NL3w15 & 0.015  &  4.9652 & 508.240  & 7.890&   5.465   &  0.238&272.25   \\
NL3w30 & 0.030  &  5.6642 & 508.270  & 7.890&   5.475   &  0.195&272.56   \\
NL3w50 & 0.050  &  7.3236 & 508.270  & 7.890&   5.496   &  0.132&272.56   \\
       \hline
FSUGw15& 0.015  &  5.0403 & 491.490  & 7.883&   5.463   &  0.248 &229.96  \\
FSUGold& 0.030  &  5.8837 & 491.500  & 7.883&   5.473   &  0.207 &230.00   \\
FSUGw45& 0.045  &  7.3695 & 491.480  & 7.883&   5.488   &  0.158 &229.92   \\
  \hline
     \end{tabular}
  \end{center}
 \end{table}

\begin{figure}[thb]
\begin{center}
\vspace*{-20mm}
\includegraphics[height=12.0cm,width=12.0cm]{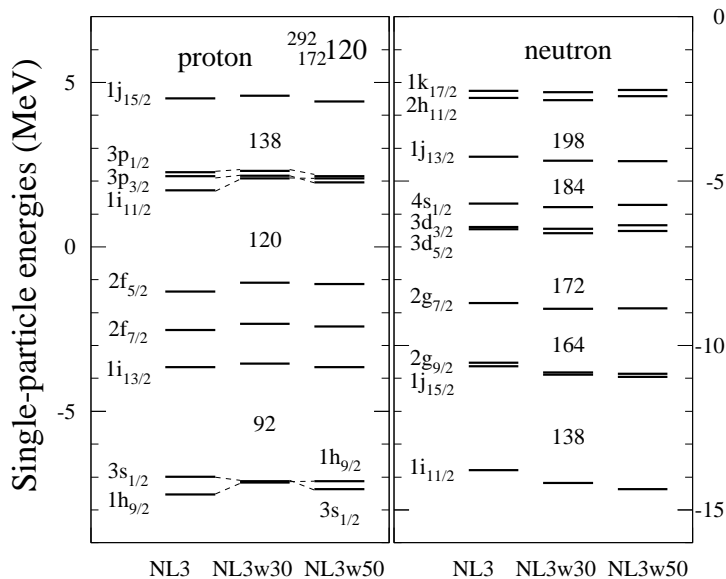}
 \end{center}
\vspace*{-10mm} \caption{Single-particle energies in $^{292}120$ with
various parameter sets in the NL3 calculations. \label{fel}}
\end{figure}
In Fig.~\ref{fel}, we plot the single-particle energies for
$^{292}120$ in the NL3 calculations. Results with various
isoscalar-isovector couplings are displayed in columns. The large
gaps for N=172 and Z=120, as shown in Fig.~\ref{fel} indicate that
the nucleus $^{292}120$ is doubly magic. It is seen that the shell
closure at N=172 and Z=120 undergoes a small but favorable
enhancement due to the inclusion of the isoscalar-isovector coupling.
As shown in the left panel, the position of $\pi1h_{9/2}$ relative to
that of $\pi3s_{1/2}$ shifts appreciably with the inclusion of the
isoscalar-isovector coupling. For the large $\Ld_{\rm v}$, even the
level inversion  takes place. This shift is in  favorably agreement
with the one called as the {\it empirical shift} in
Refs.~\cite{af03,af06}. Up to now, there is no direct data of the
single-particle energies of SHN, while the so-called empirical shift
is obtained by extracting available single-particle energies in
deformed nuclei of the $A\sim250$ region (e.g.,
$^{249}$Bk)~\cite{ah71,ah76,ah90,ah00}. Since several deformed
single-particle levels observed in the $A\sim250$ nuclei emerge from
spherical subshells in SHN, the appropriate description of empirical
shifts can provide a favorable support for the predictions on
properties of SHN, especially the nuclear magicity. With the
inclusion of the isoscalar-isovector coupling, the empirical shift
between the $\pi1h_{9/2}$ and $\pi3s_{1/2}$  can be well reproduced.
It is also interesting to see that the low-$j$ levels $\pi3p_{3/2}$
and $\pi3p_{1/2}$ can be significantly modified by the
isoscalar-isovector coupling. However, its influence on the shell
closure at Z=120 remains small. As a result, the shell gap for Z=120
is just weakly affected by the empirical shift. This means that these
parameter sets including the original NL3 can provide a reliable
prediction on the shell closure at Z=120. On the other hand, the
empirical shift for $\nu1i_{11/2}$~\cite{af03} in the single-neutron
spectrum is not reproduced with the inclusion of the
isoscalar-isovector coupling. In Ref.~\cite{af03}, it can be seen
that the shell closure at N=172 is just moderately affected by the
empirical shift.  The present prediction on the large gap for N=172
does not contradict with the early analysis with the empirical shift.
Indeed, the relative energies between $\nu2g_{9/2}$, $\nu2g_{7/2}$
and $\nu3d_{5/2}$, one of which determines the N=172 gap, are almost
independent of RMF parametrizations, see Ref.~\cite{af03} and
references therein. The situation of the neutron shell closure at
N=184 is less known,  since there is no empirical constraint on the
$\nu4s_{1/2}$. However, the N=184 gap is almost independent of the
shifts of the interior levels caused by the inclusion of the
isoscalar-isovector coupling. Similarly, the N=184 gap would not be
much affected even if the empirical shift for $\nu1i_{11/2}$ is
accurately reproduced. In this sense, the occurrence of the shell
closure at N=184 seems unlikely.

\begin{figure}[thb]
\begin{center}
\vspace*{-20mm}
\includegraphics[height=14.0cm,width=12.0cm]{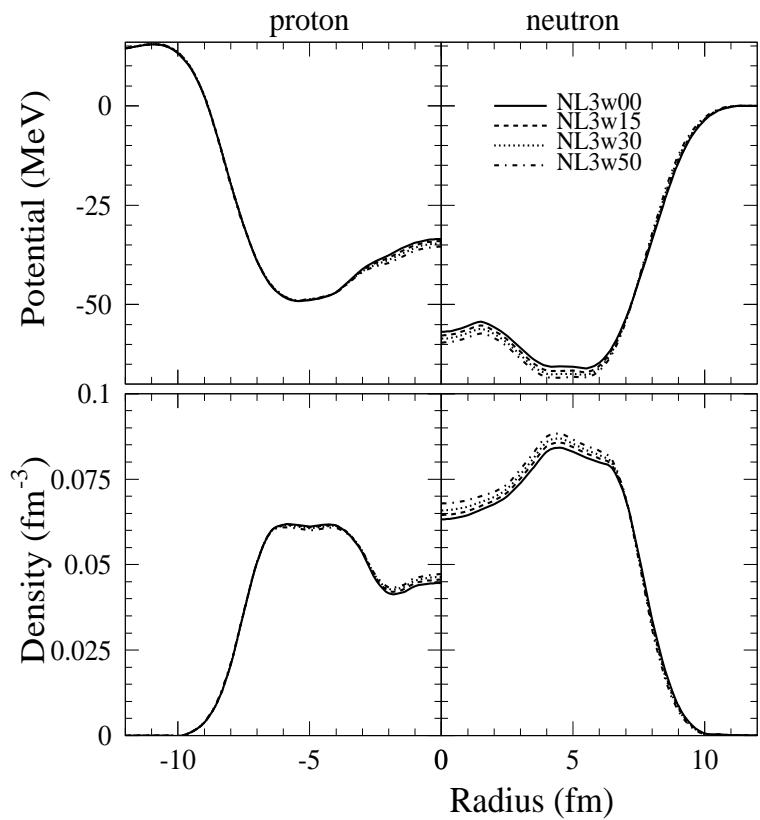}
 \end{center}
\vspace*{-10mm} \caption{Nucleon potentials (upper panels) and
nucleon density distributions (lower panels) in $^{292}120$ with
various parameter sets in the NL3 calculations. The nucleon potential
is defined as $U=V(r)-g_\sg\sg(r)$, also see
Eq.(\ref{eqpot}).\label{fpotr}}
\end{figure}
Now, let's understand  underlying factors that intrigue the
significant shift in  single-particle spectra with the inclusion of
the isoscalar-isovector coupling. As shown in Fig.~\ref{fel}, the
spin-orbit splitting can be modified by the isoscalar-isovector
coupling, and the modification increases moderately with the angular
momentum. However, the modification to the spin-orbit coupling is
just moderate and is not sufficient to cause the empirical shift.
Another factor that affects the modification to the single-particle
spectrum is the orbit-orbit interaction. The orbit-orbit interaction
can generally be given in a form of the centrifugal force, and it
reflects the flatness of the nuclear potential. In Fig.~\ref{fpotr},
the nucleon potentials and density distributions in $^{292}120$ are
plotted. As shown in the upper left panel of Fig.~\ref{fpotr}, the
homogeneity of the proton potential in the central region can be
modified by the isoscalar-isovector coupling. This modification can
bring about the moderate change in the  single-proton levels. Both
changes in the orbit-orbit and spin-orbit interactions thus lead to
significant empirical shifts of proton levels, observed in
$^{292}120$. Similarly, the shifts in the single-neutron levels can
be understood by the modification in the orbit-orbit and spin-orbit
interactions caused by the isoscalar-isovector coupling. Though the
isoscalar-isovector coupling can affect shifts in the single-particle
levels for both protons and neutrons significantly,its modification
to the proton or charge radius is much less than that to the neutron
radius. This can be seen in Fig.~\ref{fpotr} by comparing the
modification to the neutron density distribution with that to the
proton one.

\begin{figure}[thb]
\begin{center}
\vspace*{-25mm}
\includegraphics[height=12.0cm,width=12.0cm]{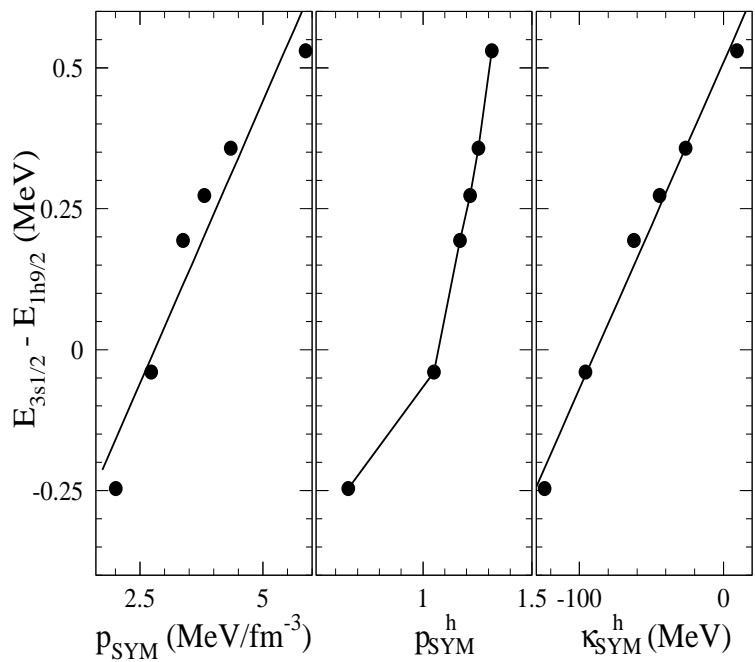}
 \end{center}
\vspace*{-10mm}\caption{Relative shift between $\pi1h_{9/2}$ and
$\pi3s_{1/2}$ as a function of the symmetry pressure and curvature.
The line in the left and right panels are obtained with a linear fit.
The pressure in the left panel is evaluated at saturation density,
while the pressure and curvature in the middle and right panels,
respectively, are calculated at half saturation density. \label{fpk}}
\end{figure}

Furthermore, it is significant to establish a quantitative
correlation between the density dependence of the symmetry energy and
the relative shift of single-particle energies. It is known that the
symmetry energy can be expanded in the vicinity of saturation density
in the following form~\cite{li98,furn02}:
\begin{equation}\label{eqsym1}
    E_{sym}(\rho_B)=E_{sym}(\r_0)+\frac{L}{3}\frac{\r_B-\r_0}{\r_0}
    +\frac{\kappa_{sym}}{18}\frac{(\r_B-\r_0)^2}{\r_0^2}+\cdots,
\end{equation}
where $L$ and $\kappa_{sym}$ are the slope and curvature of the
symmetry energy at saturation density, respectively, defined as
\begin{equation}\label{eqslp}
L=3\r_0\left.\frac{\pp E_{sym}}{\pp\r_B}\right
|_{\r_0},~~\kappa_{sym}=\left. 9\r_0^2\frac{\pp^2
E_{sym}}{\pp\r_B^2}\right |_{\r_0}.
\end{equation}
The slope of the symmetry energy defines the symmetry pressure
through the relation $p_{sym}=\r_0L/3$. In the history, the proper
inclusion of the spin-orbit coupling played an important role in
giving rise to the correct shape of nuclear potential and hence the
ordering of the energy levels. Since the spin-orbit interaction is
associated  with the surface property subject to the subsaturation
density region, it is useful to similarly define the slope and
curvature of the symmetry energy at half saturation density, $L^h$
$(p_{sym}^h)$ and $\kappa^h_{sym}$. As seen in Fig.~\ref{fel}, the
relative shifts of the orbitals change with respect to the
isoscalar-isovector coupling. Since the correlation between these
relative shifts and the density dependence of the symmetry energy are
similar,  we just plot as an example in Fig.~\ref{fpk} the relative
shift between $\pi1h_{9/2}$ and $\pi3s_{1/2}$ as a function of the
symmetry pressure and curvature. As shown in Fig.~\ref{fpk}, the
relative shift of these two levels is approximately linear in the
symmetry pressure (at saturation density) and correlates
quadratically with the symmetry pressure at half saturation density.
A stronger correlation at half saturation density reflects the strong
dependence of single-particle energies on the surface property of
finite nuclei. In the right panel, it is shown that the relative
shift of the two levels is linear in the curvature at half saturation
density, which is consistent with the quadric correlation as shown in
the middle panel. It was found in the early time for the RMF theory
that the proper density dependence of the potential is important for
a correct spin-orbit potential and hence the ordering of orbitals,
e.g. see~\cite{bo77,bo83}. Similarly, it is here interesting to see
that the density dependence of the symmetry energy (or of the
isovector potential) affects moderately the single-particle energies.
In particular, in the present work we can obtain the relative shift
between $\pi1h_{9/2}$ and $\pi3s_{1/2}$ with the appropriate density
dependence of the symmetry energy. The correlations shown in
Fig.~\ref{fpk} also exist for other SHN, even if the central
depression disappears. Here, the central depression plays a role in
enhancing the correlation strength ( e.g. the slope in the linear
correlation).

\begin{figure}[thb]
\begin{center}
\vspace*{-25mm}
\includegraphics[height=14.0cm,width=12.0cm]{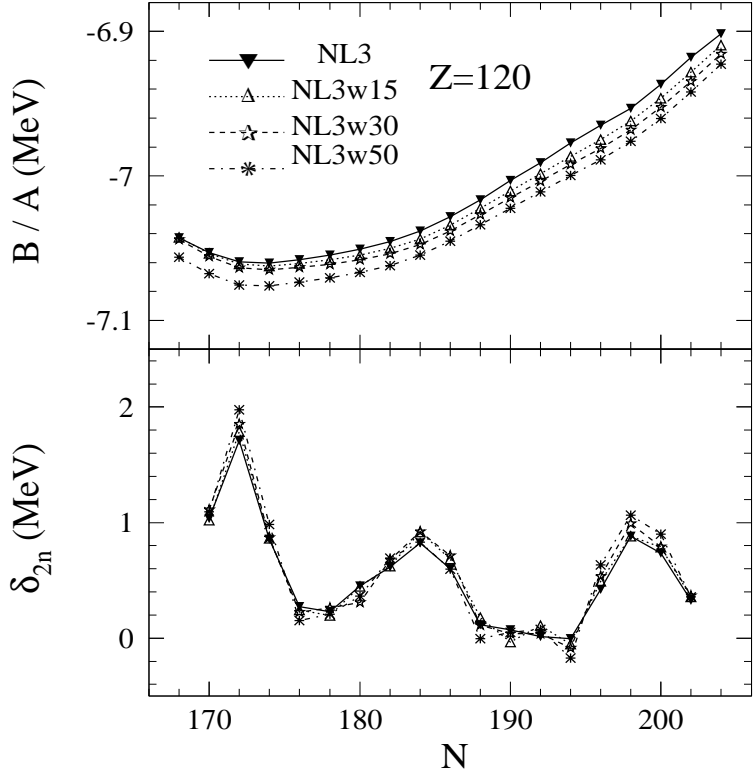}
 \end{center}
\vspace*{-10mm}\caption{Binding energies per nucleon of Z=120
isotopes (upper panel) and two-neutron gaps (lower panel) with
various parameter sets in the NL3 calculations.\label{fbnl3}}
\end{figure}
Next, we turn to the discussion on the two-nucleon gaps. Besides the
gap in the single-particle spectrum, a direct measure of the shell
closure is the appearance of the peak in the two-nucleon gaps, which
are defined as~\cite{rut97}:
\begin{eqnarray}
    \dt_{2n} &=& S_{2n}(N+2,Z)-S_{2n}(N,Z)=2B(N,Z)-B(N-2,Z)-B(N+2,Z), \\
  \dt_{2p} &=&S_{2p}(N,Z+2)-S_{2p}(N,Z)= 2B(N,Z)-B(N,Z-2)-B(N,Z+2),
\end{eqnarray}
where $S_{2n}$ and $S_{2p}$ are the two-neutron and two-proton
separation energies, respectively. The peak of the two-nucleon gap
reflects the large change of the two-nucleon separation energy,
signaling the shell closure. Moreover, the two-nucleon gap can
reflect appropriately the gap size  in the single-particle
spectrum~\cite{ben99}. In Fig.~\ref{fbnl3}, we plot the binding
energy per nucleon (upper panel) and the two-neutron gap (lower
panel) for the Z=120 isotopes in the NL3 calculations. As shown in
the upper panel of Fig.~\ref{fbnl3}, the difference between binding
energies is small for different isoscalar-isovector couplings, and
this is attributed to the refitting of $m_\sg$.  As shown in the
lower panel of Fig.~\ref{fbnl3},  the two-neutron gap at N=172 can
earn a moderate rise with the inclusion of the isoscalar-isovector
coupling, which is consistent with the observation of the
single-neutron spectrum in Fig.~\ref{fel}. The similar increasing
tendency also occurs for the N=198 shell gap.  As shown in the lower
panel of Fig.~\ref{fbnl3}, the peak of the two-neutron gap occurs at
N=172, 184 and 198 in the Z=120 isotopes. Comparing to the sharp peak
at N=172, peaks at N=184 and 198 becomes much blunt. This indicates
that the shell gaps at N=184 and 198 are not well developed.  Also,
we examine the $\dt_{2n}$ for the Z=126 isotopes, and the case is
similar.

As shown in Fig.~\ref{fbnl3}, the two-neutron gap can be affected by
the isoscalar-isovector coupling. In some isotopes such as
$^{292}120$ and $^{318}120$, the two-neutron gap can gain a rise with
the inclusion of the isoscalar-isovector coupling. However, the
modification to the two-neutron gap caused by the isoscalar-isovector
coupling is not governed by the isotopic effect in SHN. For instance,
this is clear as comparing the modification to $\dt_{2n}$ at N=172
with the one at N=184. In fact, the relatively pronounced
modification is associated with the specific geometries such as the
central depression or enhancement. It was pointed out in
Ref.~\cite{af05} that the central depression in $^{292}120$ is
predominantly from the proton occupation of high-j orbitals, while
the neutron central depression results from the strong coupling
between protons and neutrons. In presence of the central depression,
the change in the neutron potential and density distribution caused
by the isoscalar-isovector coupling exhibits a radial inhomogeneity,
as shown in right panels of Fig.~\ref{fpotr}. This inhomogeneity
causes consistently  modifications to the level shifts and the
two-neutron gaps. With the increase of the neutron numbers, the
neutron central depression tends to disappear, while the central
enhancement appears with more neutrons. The inhomogeneity of the
modifications produced in presence of the central enhancement
explains the change in the two-neutron gap in more neutron-rich
isotopes, as shown in the lower panel of Fig.~\ref{fbnl3}.

\begin{figure}[thb]
\begin{center}
\vspace*{-25mm}
\includegraphics[height=12.0cm,width=12.0cm]{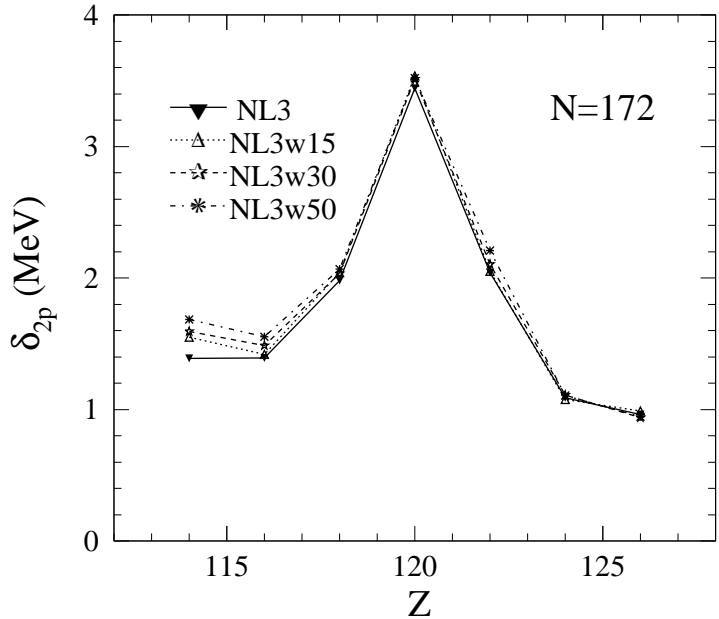}
 \end{center}
\vspace*{-10mm}\caption{Two-proton gap $\dt_{2p}$ for N=172 isotones
with various parameter sets in the NL3 calculations. \label{fznl3}}
\end{figure}
To examine the proton shell closure, we plot in Fig.~\ref{fznl3} the
two-proton gap for the N=172 isotones. As shown in Fig.~\ref{fznl3},
there is only one peak at Z=120. No peak is observed at Z=114 and
126. The proton shell closure at Z=126 was predicted by the
Hartree-Fock approach with a variaty of Skyrme
interactions~\cite{rut97,ben99}, while the RMF models disincline the
appearance of this shell closure. As seen in Fig.~\ref{fel}, though
the isoscalar-isovector coupling tends to shift the relative position
between $\pi1i_{11/2}$ and $\pi3p$ orbitals,  the formation of the
Z=126 shell closure does not appear. In RMF models, the Z=114 shell
closure was only predicted by parameter sets NL-SH and NL-RA1 due to
the relatively large spin-orbit splitting of $2f$
orbitals~\cite{la96,ra01}. In NL3, the spin-orbit splitting of $2f$
orbitals is not large, and though the modification of the
isoscalar-isovector coupling to $\dt_{2p}$ is rather pronounced, as
shown in Fig.~\ref{fznl3}, it is far from sufficient to form the
Z=114 shell gap. It is necessary to point out that the isotopic
effect is  prominent for the two-proton gap. For instance, as
observed in Fig.~\ref{fznl3}, the sensitivity of the $\dt_{2p}$ to
the isoscalar-isovector coupling differs clearly for the N=172
isotones with Z=114 and 116. Indeed, the pronounced isotopic effect
exists for the Z=120 shell closure. $\dt_{2p}=3.5$ MeV in $^{292}120$
reduces to 2.7 MeV in $^{304}120$. Further, the peak at Z=120
disappears totally in $^{318}120$.  As far as the double shell
closure in spherical SHN is concerned, $^{292}120$ turns out to be
the most possible candidate with various parameter sets in the NL3
calculations. This is consistent with the prediction in
Refs.~\cite{ben99,rut97,zh05,af05}.

\begin{figure}[thb]
\begin{center}
\vspace*{-20mm}
\includegraphics[height=12.0cm,width=12.0cm]{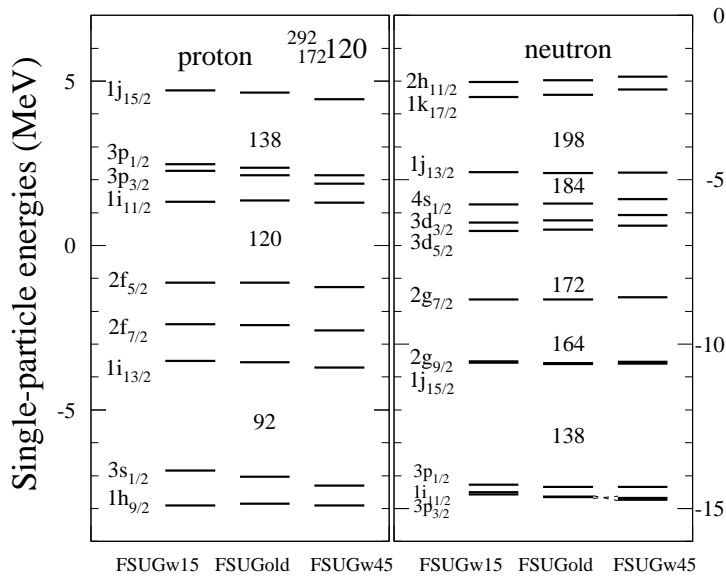}
 \end{center}
\vspace*{-10mm} \caption{Single-particle energies in $^{292}120$ with
various parameter sets in the  FSUGold calculations. \label{felf}}
\end{figure}

Now, we discuss the results with the FSUGold. In Fig.~\ref{felf}, the
single-particle energies for $^{292}120$ are plotted with various
parameter sets in the  FSUGold calculations.  Though the
isoscalar-isovector coupling is already included in the FSUGold, its
strength is changed in FSUGw15 and FSUGw45 in order to manifest the
importance of this coupling. Compared to results shown in
Fig.~\ref{fel}, the role of the isoscalar-isovector coupling in the
single-particle properties is similar in FSUGold, and here the
nucleus $^{292}120$ is also doubly magic. Though the role of the
isoscalar-isovector coupling is less prominent for the empirical
shift between $\pi3s_{1/2}$ and $\pi1h_{9/2}$ than that in the NL3,
it can favorably reduce the  relative energy between these two
levels. Similarly,  we can establish correlations as in
Fig.~\ref{fpk} for the FSUGold results, while here for brevity we
neglect the display of the correlations. On the other hand, some
distinctions of the single-particle spectrum are given with the
FSUGold. For instance, an inversion of levels $\nu1k_{17/2}$ and
$\nu2h_{11/2}$ is observed in the FSUGold calculations, while it does
not take place in the NL3 calculations. The similar inversion also
occurs between the $\nu1i_{11/2}$ and $\nu3p$ orbitals. Moreover, the
shell gap at N=184 is suppressed as compared to that with the NL3. On
the contrary, the gap for N=198 in FSUGold well develops, and the
isoscalar-isovector coupling can further enhance the magnitude of
this gap. This implies that N=198 is a magic number with the FSUGold.
With the moderate increase of the proton number, the N=198 shell gap
remains large. For instance, the size of the N=198 shell gap in
$^{324}126$ is even a little larger than that in $^{318}120$.

\begin{figure}[thb]
\begin{center}
\vspace*{-25mm}
\includegraphics[height=14.0cm,width=12.0cm]{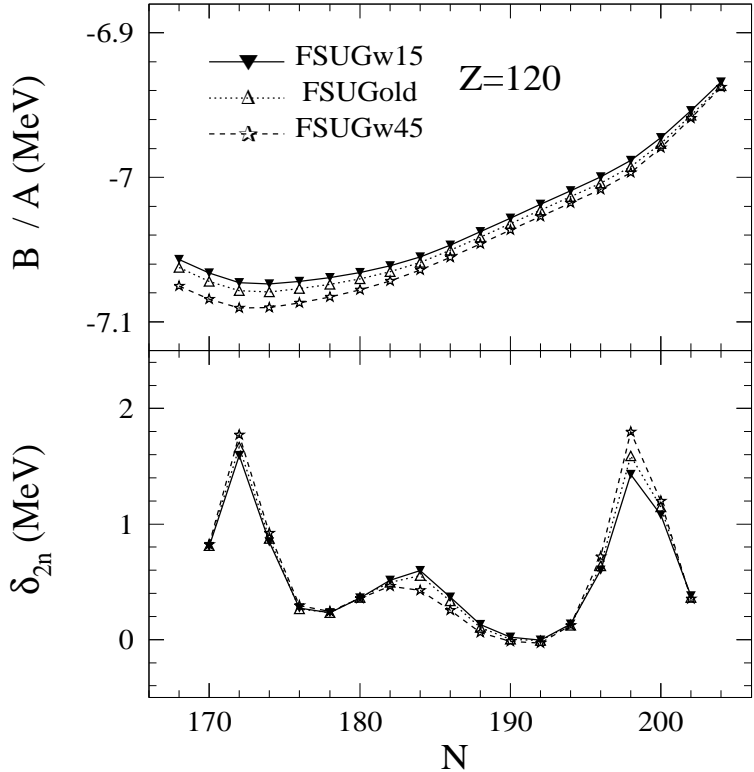}
 \end{center}
\vspace*{-10mm}\caption{The same as in Fig.~\ref{fbnl3} but for the
FSUGold calculations.\label{fbfsu} }
\end{figure}
The N=198 shell closure can consistently be observed using the
$\dt_{2n}$. Fig.~\ref{fbfsu} displays the binding energies and
two-neutron gaps for the Z=120 isotopes. As shown in the lower panel
of Fig.~\ref{fbfsu}, the two-neutron gap at N=198 is clearly
increased by softening the symmetry energy. Besides the large N=172
gap, we see that the large N=198 gap emerges. The situation of the
shell closure at N=198 in FSUGold differs from that in NL3. This
distinction can be associated with different model constructions. For
instance, the nonlinear self-interaction of the $\om$ meson is
included in FSUGold. Moreover, the compression modula of the NL3 and
FSUGold differ by about 40 MeV. In general, the lower
incompressibility in FSUGold allows more nucleons to be accommodated
in the potential well.  For the two-proton gap in FSUGold, it is less
sensitive to differences in the symmetry energy than that in NL3,
while the isotopic effect is similar to that in NL3.  The increase of
the neutron number suppresses the Z=120 proton gap. The two-proton
gap is 2.7 and 2.3 MeV with N=172 and 184, respectively. As compared
to the NL3 results, the value of $\dt_{2p}$ with the FSUGold turns
out to be smaller. The peak disappears with N=198, similar to that
with the NL3.

\begin{figure}[thb]
\begin{center}
\vspace*{-20mm}
\includegraphics[height=18.0cm,width=12.0cm]{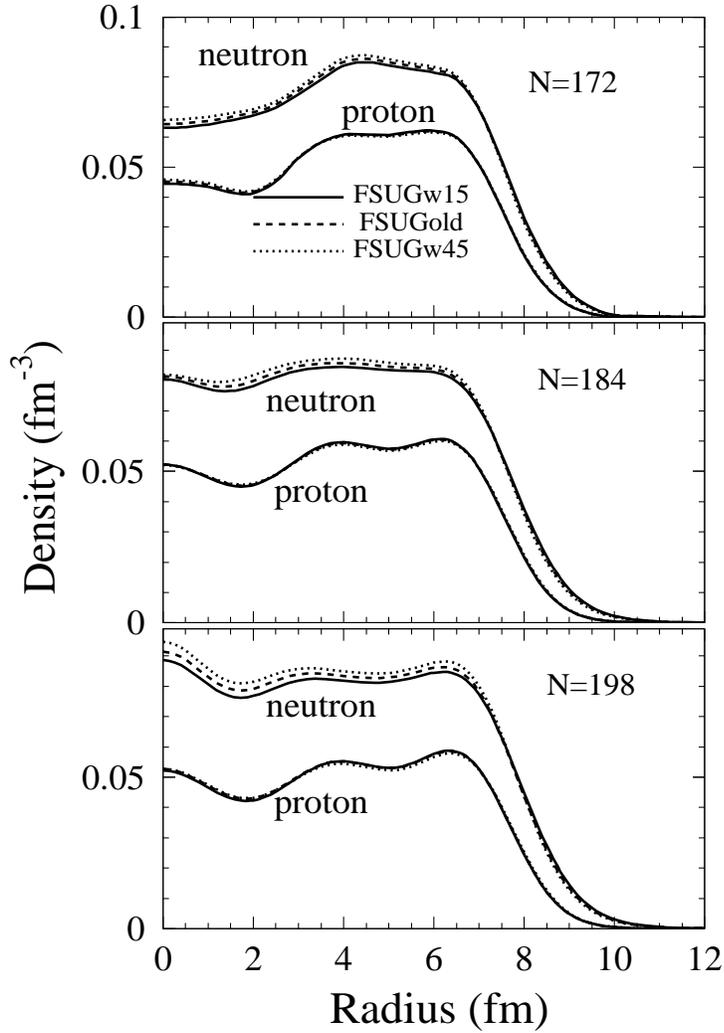}
 \end{center}
\vspace*{-10mm} \caption{Nucleon density distributions in the Z=120
isotopes with various parameter sets in the  FSUGold calculations.
\label{fdfsu}}
\end{figure}

To examine the consistency of the calculation,  we plot in
Fig.~\ref{fdfsu} the change in nucleon density distributions in the
FSUGold calculations. It is shown that in $^{318}120$ the neutron
density is enhanced in the central region. In presence of this
central enhancement, the radial inhomogeneity appears to increase the
sensitivity of the two-neutron gap to differences in the symmetry
energy.  For $^{292}120$, it was mentioned in Ref.~\cite{af05} that
the magnitude of the central depression increases with the decrease
of the compression modulus. We note that the central depression in
$^{292}120$ with the FSUGold is not more prominent than that with the
NL3 as shown in Fig.\ref{fpotr}. Indeed, this can be attributed to
the non-linear $\om$-meson self-interaction in FSUGold that lowers
the potential barrier.   For $^{304}120$, as shown in the middle
panel of Fig.~\ref{fdfsu}, there exists an inhomogeneity of the
modification added by the isoscalar-isovector coupling,  and this is
consistent with the moderate decrease of the two-neutron gap at N=184
with the $\Ld_{\rm v}$, as seen in the lower panel of
Fig.~\ref{fbfsu}.

\begin{figure}[thb]
\begin{center}
\vspace*{-20mm}
\includegraphics[height=16.0cm,width=12.0cm]{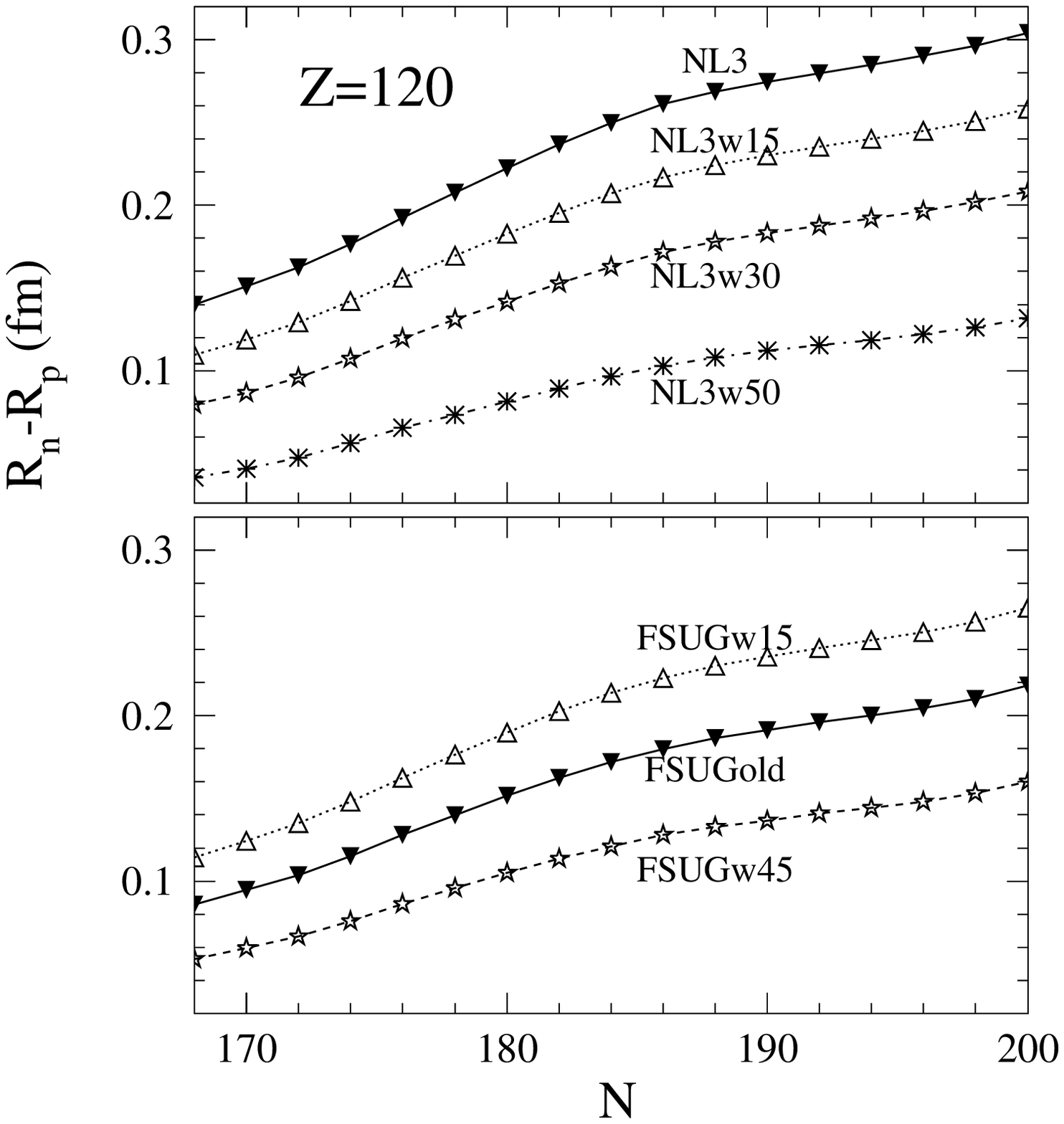}
 \end{center}
\vspace*{-10mm} \caption{Neutron skin thicknesses for SHN with
various parameter sets in the NL3  (upper panel) and FSUGold (lower
pannel) calculations. \label{fnsnf}}
\end{figure}
Next, we investigate the neutron skins in SHN. As it is known, the
neutron skin thickness in heavy nuclei such as $^{208}$Pb is
sensitive to differences in the symmetry energy.  In SHN, the case is
similar. In Fig.~\ref{fnsnf}, we display neutron skin thicknesses in
Z=120 isotopes  for various parameter sets  in the NL3 and FSUGold
calculations.  In general, the sensitivity of the neutron skin
thickness to differences in the symmetry energy can be well
understood in the following way~\cite{brown00,furn02,chen05}. The
pressure of pure neutron matter at saturation density is equal to the
symmetry pressure. As the symmetry pressure decreases with the
inclusion of the isoscalar-isovector coupling, the neutron skin
thickness reduces in neutron-rich SHN.  On the other hand, two
interesting features in Fig.~\ref{fnsnf} are also observed. First,
the neutron skin thicknesses in various SHN can roughly be separated
into two reaches according to the slope. The reach with the larger
slope is associated with the large neutron gap at N=172. With the
addition of neutrons above this gap the neutron skin thickness thus
increases clearly. As the occupation surpasses the much smaller gap
at N=184, the nuclear attraction from the interior gap can still
appreciably restrain the extension of neutrons. This leads to a
smaller slope at the larger isospin asymmetry. Second, we observe
that the difference of the neutron skin thickness increases
moderately at large isospin asymmetries.  Because of the looser
binding with the increase of the isospin asymmetry, it gives rise to
an enhanced sensitivity to differences in the symmetry energy.
Indeed,  the correlation between the neutron skin and  symmetry
pressure in SHN deviates from a simple linear relation for heavy
nuclei due to the more extended neutron distribution given by a
looser binding compared to that of heavy nuclei.

\begin{figure}[thb]
\begin{center}
\vspace*{-20mm}
\includegraphics[height=16.0cm,width=12.0cm]{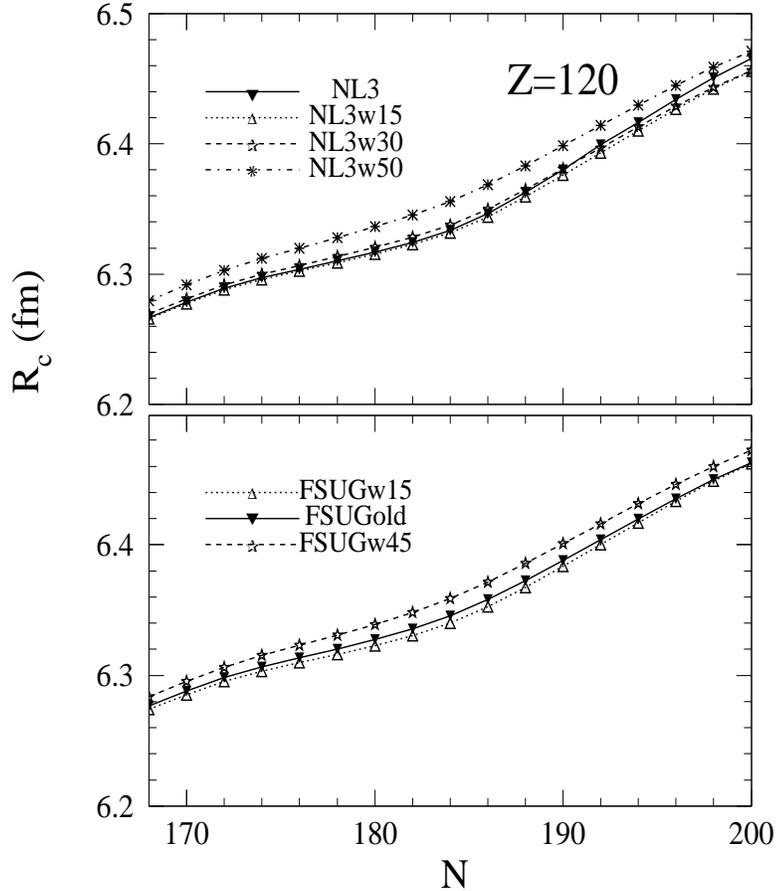}
 \end{center}
\vspace*{-10mm} \caption{The same as in Fig.~\ref{fnsnf} but for
charge radii. \label{fpnf}}
\end{figure}
It is necessary to mention that the uncertainty of the neutron skin
thickness is predominately from the changes in the neutron radius.
Fig.~\ref{fpnf} displays the charge radii of the Z=120 isotopes for
various parameter sets in the NL3 and FSUGold calculations. It is
shown that the change in charge (or, proton) radii is much smaller
than the corresponding neutron skin thickness that is shown in
Fig.~\ref{fnsnf}. Moreover, for a considerable domain, it is seen
that the charge radius of SHN can be well approximated as a constant.
This is favorable, as one expects that the inclusion of the
isoscalar-isovector coupling shouldn't compromise the success of
models in reproducing a variety of ground-state properties. Recently,
the precision measurement of the neutron radius of $^{208}$Pb has
been proposed at the Jefferson laboratory via the parity-violating
electron scattering on neutrons in $^{208}$Pb~\cite{jeff}. The
measurement of the neutron radius that promises 1\% accuracy will
impose a strict constraint on the density dependence of the symmetry
energy. Correspondingly, this also provides a significant constraint
on the properties of SHN through the relationship that can be
established for properties between SHN and $^{208}$Pb.

At last, we mention a few weak points in this work. First, with the
isoscalar-isovector coupling included or changed in the best-fit
models, we fitted the corresponding parameters  without using the
best-fit procedure. Considering the important effect of the
isoscalar-isovector coupling on the empirical shift, its inclusion
seems necessary in the construction of the best-fit models for the
study of SHN in future.  Second, we note that the effect of the
isoscalar-isovector coupling on the empirical shift of neutron levels
is not as satisfactory as that of proton levels, though the
prediction on the N=172 shell closure in SHN is not much affected by
this coupling. For a more detailed investigation of level shifts in
future, it may be favorable to consider the coupling with the surface
vibration modes~\cite{mah85,vre02} and the influence of the
relatively small Lorentz mass of nucleons in RMF models. Third,  in
this work we have used the simple BCS theory with phenomenological
pairing gaps. For this point, we address it in some detail. It would
be interesting to treat the pairing interaction in a dynamical way
such as in the relativistic Hartree-Boguliubov (RHB) theory (for
reviews, e.g., see Refs.~\cite{af03,ri96,meng06}) for the open-shell
SHN that are usually deformed, e.g., see
Refs.~\cite{ren01,ren02,ren021}, though the effect of pairing
interactions on the shell closure is small. To our knowledge,  a RHB
model in the coordinate space is still not available for deformed
nuclei, especially SHN~\cite{meng06}, and in the RHB model the
calculation for deformed nuclei is usually performed with the
harmonic oscillator basis expansion. Due to the numerical
complication, the RHB calculation in the deformed framework is still
limited, see Refs.~\cite{af06,tian09} and references therein. In
fact, most works treat the nucleon pairings in open-shell and
deformed SHN using the BCS theory. It is interesting to note that the
RMF model plus the BCS pairing works quite well for the open-shell
nuclei except drip-line nuclei~\cite{lijq02,jiang05}, since for most
isotopes the nucleon occupation number in the continuum is just
moderate. For most open-shell SHN, we find that the occupation in the
continuum is also comparatively small. In this sense, the results
obtained for most open-shell SHN with the RMF model plus the BCS
pairing are comparable to those obtained with the RHB model.
Recently, a separable pairing force was proposed for the RHB model
with considerable reduction of the computing time~\cite{tian09}, and
it would be hopefully developed to study properties of open-shell
SHN.

\section{Summary}
\label{summary}

In summary, we have investigated the dependence of the ground-state
properties of spherical SHN on the density dependence of the symmetry
energy within RMF models. The various density dependences of the
symmetry energy are simulated by changing the strength of the
isoscalar-isovector coupling in RMF models (the NL3 and FSUGold). It
is found that the isoscalar-isovector coupling produces an important
effect on the empirical shift of spherical orbitals in SHN.
Especially, the empirical shift between the $\pi3s_{1/2}$ and
$\pi1h_{9/2}$ in NL3 is nicely reproduced by including the
isoscalar-isovector coupling. This provides a favorable support for
the Z=120 shell closure. In addition, the isoscalar-isovector
coupling can produce a small but favorable effect on the N=172 shell
closure.   With both models NL3 and FSUGold that differ in the
compression modulus, the double shell closure is predicted in
$^{292}_{172}120$. The shell closure is also investigated in more
Z=120 isotopes. We have discussed the association of the central
depression or enhancement with the effect of the softening of the
symmetry energy in SHN. Compared to the moderate modification to
single-particle energies and  shell gaps, significant reduction of
the neutron skin thickness in SHN is expectedly obtained by softening
the symmetry energy. Moreover, the proton radius is little changed,
similar to the situation in $^{208}$Pb.

\section*{Acknowledgement}

The work was supported in part by the National Natural Science
Foundation of China under Grant No. 10975033, the China Jiangsu
Provincial Natural Science Foundation  under Grant No.BK2009261, the
Knowledge Innovation Project of the Chinese Academy of Sciences under
Grant No. KJXC3-SYW-N2, and the China Major State Basic Research
Development Program under Contract No. 2007CB815004.

\end{document}